\documentclass[osajnl2,twocolumn,showpacs,superscriptaddress]{revtex4}

\usepackage{amsmath}
\usepackage{amssymb}

\newcommand{\be}{\begin{equation}}
\newcommand{\ee}{\end{equation}}
\newcommand{\bea}{\begin{eqnarray}}
\newcommand{\eea}{\end{eqnarray}}

\usepackage{here}
\usepackage{graphicx}% Include figure files
\graphicspath{{pictOL/}{pict/}{}}

\usepackage{dcolumn}% Align table columns on decimal point
\usepackage{bm}% bold math

\newcommand\pictc[5]{\begin{figure}[t]
                       \centerline{\vspace{-0mm}
                        \includegraphics[width=#1\columnwidth,height=0.7\textheight,keepaspectratio]{#3}}
                       \protect\caption{\protect\label{fig:#4} #5}\vspace{-0mm}
                    \end{figure}            }

\newcommand\pict[4][1]{\pictc{#1}{!tb}{#2}{#3}{#4}}

\newcommand\rpict[1]{\ref{fig:#1}}

\newcommand\leqt[1]{\protect\label{eq:#1}}
\newcommand\reqtn[1]{\ref{eq:#1}}
\newcommand\reqt[1]{(\reqtn{#1})}

\newcounter{Fig}

\begin{document}

\title{All-optical switching of dark states in nonlinear coupled microring resonators}

\author{Jacob Scheuer$^{*}$}
\affiliation{School of Electrical Engineering, Department of Physical Electronics, Tel-Aviv University, Ramat-Aviv, Israel}

\author{Andrey A. Sukhorukov}
\author{Yuri S. Kivshar}
\affiliation{Nonlinear Physics Centre, Research School of Physics and Engineering, Australian National University, Canberra ACT 0200, Australia\\
$^{*}$ Corresponding author: kobys@eng.tau.ac.il}

\begin{abstract}
We propose and analyze an on-chip all-optical dynamical tuning scheme for coupled nonlinear resonators employing a single control beam injected in parallel with a signal beam. We show that nonlinear Kerr response can be used to dynamically switch the spectral properties between ``dark-state'' and electromagnetically-induced transparency configurations. Such scheme can be realized in integrated optical applications for pulse trapping and delaying.
\end{abstract}

\ocis{\small (190.5940) Self-action effects;
             (190.0190) Nonlinear optics
        }

\maketitle

New approaches for optical pulse control in photonic structures employ an analogy with the coherence phenomena in quantum-mechanical systems. In particular, electromagnetically-induced transparency (EIT) is attributed to destructive quantum interference with a narrow transparent window in the absorption spectrum.  In optics, coherent interference of the resonant modes of coupled optical cavities can exhibit EIT-like transmission and QM-like phenomena at room temperature~\cite{Xu:2006-123901:PRL}, thus relaxing the bandwidth and de-coherence constraints of QM. Recently, several studies have shown that dynamically tuning a system of two coupled micro-ring into and out of a ``dark-state''can facilitate all-optical trapping and releasing of optical pulses~\cite{Xu:2007-406:NPHYS, Dong:2009-2315:OL}.

Various experimental approaches for tuning the resonator frequencies have been developed to suit particular material platforms. In Si structures, out-of-plane optical pump was used for carrier excitation~\cite{Xu:2007-406:NPHYS, Dong:2009-2315:OL} or thermal heating~\cite{Xu:2006-6463:OE}. Electro-optic control with integrated p-i-n junctions~\cite{Manipatruni:2008-1644:OL} and metal film heaters~\cite{Sherwood-Droz:2008-15915:OE}  have also been demonstrated. In AlGaAs structures~\cite{Van:2002-705:ISQE}, thermal and two-photon absorption effects have been utilized for optical signal processing. However, it remains an open question how to realize fully all-optical scheme for dynamic switching of ``dark-state'' in coupled microring resonators, where the signal and control pulses can propagate within the photonic chip.

In this Letter we present a new all-optical approach for tuning coupled microring resonators with Kerr nonlinearity, using
a strong control beam launched into the photonic structure in parallel with the signal beam.
In particular, we demonstrate switching the spectral properties between a ``dark-state'' and EIT-like states. The realization of such tunability opens a way for fully on-chip trapping, releasing, and delaying of signal optical pulses.
Our approach can be realized, in particular, with the chalcogenide glass platform~\cite{Taeed:2007-9205:OE} or in AlGaAs structures at the telecom spectral range~\cite{Hartsuiker:2008-83105:JAP}, whereas nonlinear losses in Si platform can also be suppressed if the pump is tuned to longer wavelengths to avoid two-photon absorption~\cite{Harding:2009-610:JOSB}.

\pict{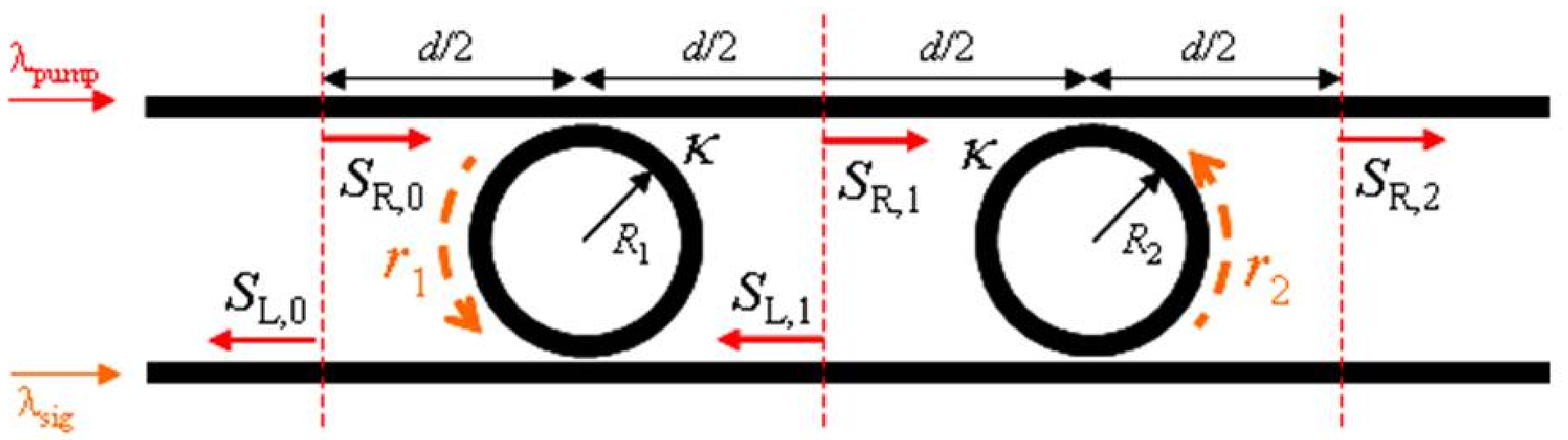}{scheme}{
(Color online)
Schematic of the coupled microrings resonators structure.
}

We consider a coupled resonator configuration consisting of two microrings side coupled to two waveguides, see Fig.~\rpict{scheme}. In general, $R_1 \ne R_2$. Although optical tuning and switching in such photonic structures has been studied~\cite{Xu:2007-406:NPHYS, Dong:2009-2315:OL}, we suggest here a different approach, where the signal and control (pump) beams are launched into the lower and upper waveguides, respectively. This scheme is also different from the configuration where the pump and signal are launched into the same waveguide~\cite{Turner:2008-4881:OE}, as the resonator-enhanced four-wave-mixing (FWM) of co-propagating waves can lead to frequency conversion. In our case, the pump and signal waves propagate in the opposite directions in both rings and waveguides, and FWM leads to effective modification of refractive index essentially without frequency conversion.

The structure is equivalent to a Fabry-Perot (FP) cavity with wavelength-dependent mirrors created by two resonators; unlike conventional FP cavities, the forward and backward propagating waves are traveling through different optical paths. When the device is set to a ``dark-state'', the structure serves as a compound, high-quality, FP resonator which can force a pulse centered at its resonance frequency to circulate inside for long periods of time.

As the pump wave propagates through the resonators, it changes the refractive index via the Kerr nonlinearity. If the losses are small, the control wave intensity is constant and the associated index changes are homogeneous inside each sections of the rings and the waveguides between the successive coupling regions. We denote by $\Delta n_{jR}$ and $\Delta n_{jL}$ the nonlinear refractive index changes in the right and left half-rings, respectively, where $j=1,2$ is the number of the resonator. Since the optical field is enhanced inside the rings and can exceed significantly the input intensity, we neglect the nonlinearity in the straight waveguides. At steady-state, when $\Delta n$ are constant, the intensities in half-rings ($I_{jL}$ and $I_{jR}$) and the transmitted intensity ($I_T$) can be found analytically:
\begin{equation} \leqt{stationary}
  \begin{array}{l} {\displaystyle
     I_{1L} = I_{\rm in} \kappa (1-\kappa)
            \left|-\kappa e^{i \varphi_{2R} + 2 i \varphi_{w}} + \rho_2 e^{i \varphi_{1R}}\right|^2
             |D|^{-2} ,
  } \\*[9pt] {\displaystyle
     I_{1R} = I_{\rm in} \kappa
            \left|-\kappa e^{i \varphi_{2R} + 2 i \varphi_{w}} + \rho_2 e^{-i \varphi_{1L}}\right|^2
            |D|^{-2} ,
  } \\*[9pt] {\displaystyle
     I_{2R} = I_{\rm in} \kappa (1-\kappa)
                 \left|1 - e^{i \varphi_{1L} + i \varphi_{1R}} \right|^2
              |D|^{-2} ,
  } \\*[9pt] {\displaystyle
     I_{2L} = I_{2R} (1-\kappa),\,
     I_{T} = I_{2R} \left|1 - e^{i \varphi_{2L} + i \varphi_{2R}} \right|^2 ,
  } \end{array}
\end{equation}
where
     $\rho_j = 1 - (1-\kappa) \exp(i \varphi_{jR} + i \varphi_{jL})$,
     $D = \rho_1 \rho_2 - \kappa^2 \exp(i \varphi_{1L} + i \varphi_{2R} + 2 i \varphi_{w})$,
$\kappa$ is the coupling constant between the waveguides and the ring resonators, $\varphi$ denote the phase accumulation at each half-rings and waveguide sections (either top or bottom) between the rings, $\varphi_{jR} = \pi R_j K (n_{j0} + \Delta n_{jR})$, $\varphi_{jL} = \pi R_j K (n_{j0} + \Delta n_{jL})$, $\varphi_{w} = d K n_{w0}$, $K = 2 * \pi / \lambda$ is the wavenumber, $\lambda$ is the wavelength in vacuum, $n_{j0}$ and $n_{w0}$ are the effective linear refractive indexes for the fundamental guided modes in the rings and the waveguides, respectively. These expressions generalize the model presented in Ref.~\cite{Xu:2006-123901:PRL} where identical parameters for the left and right halves of the microrings were considered.

Depending on the structure parameters, such as the resonance frequencies of the microrings and the FP, the compound cavity can be toggled between ``open'' and ``closed'' configuration. In order to trap an optical pulse in the compound cavity (``close'' the trap) it is necessary to formally satisfy the condition that the field inside the rings can be non-vanishing for zero input intensity.
This leads to the following phase conditions: (i)~$\varphi_{jL} + \varphi_{jR} = 2 \pi m_j$, i.e. the microrings are in resonance, and (ii)~$i \varphi_{1L} + i \varphi_{2R} + 2 i \varphi_{w} = 2 \pi m_c$, where $m_j$ and $m_c$ are integers. A cavity ``dark state'' is realized when these conditions are satisfied simultaneously.

\pict{fig02}{transm}{
(Color online)
Nonlinear cavity tuning: (a,b)~no pump, (c,d)~pump wavelength and input intensity determined through self-consistent approach to achieve ``dark state'' for signal wave.
Shown are (a,c)~linear transmission intensities and (b,d)~the corresponding signal intensities in the left half-rings of the first (dashed line) and second (solid line) resonators, relative to the input signal intensity.
}

We now demonstrate that the compound cavity can be dynamically tuned to and away from the conditions for a dark state by injecting a pump beam into the structure  (see Fig.~\rpict{scheme}) which induces a refractive index change through Kerr nonlinearity of the material composing the microrings. Unlike the out-of-plane pumping schemes~\cite{Xu:2007-406:NPHYS, Dong:2009-2315:OL}, the build up of the pump signal in the microrings enhances the nonlinear response.
We note that the field inside the microrings can be strongly enhanced at resonance for small values of the coupling coefficient $\kappa$, and in this regime the intensities in left and right half-rings are very similar according to Eq.~\reqt{stationary}. Therefore, we assume that approximately
$I_{jL}^{(p)} = I_{jR}^{(p)} \simeq I_j^{(p)}$, and
$\Delta n_{jR}^{(p)} = \Delta n_{jL}^{(p)} \simeq \Delta n_{j}^{(p)} = 2 \pi n_2 I_j^{(p)}$, where $n_2$ is a nonlinear coefficient. The index change at a signal wavelength $\lambda^{(s)}$ larger by a factor of two compared to that of the pump because of the nature of cross phase modulation (XPM)~\cite{Boyd:1992:NonlinearOptics}, $\Delta n_{jR}^{(s)} = \Delta n_{jL}^{(s)} = \Delta n_{j}^{(s)} = 4 \pi n_2 I_j^{(p)}$. Thus, the ``dark state'' conditions at the signal wavelength are satisfied when $I_j^{(p)} = [m_j / r_j - 2 \pi n_{j0} / \lambda^{(s)}] / ( 4 \pi n_2 \lambda^{(s)})$.

After determining the required pump intensities inside the microrings, we need to find the values of the experimentally controllable pump parameters, the wavelength $\lambda^{(p)}$ and input intensity $I_{\rm in}^{(p)}$.
To solve the nonlinear problem yielding the necessary wavelength and power of the pump we employ a self-consistent argument:
1.~We know the effective $\Delta n_{j}^{(p)}$ needed to achieve a ``dark state'' at the signal wavelength.
2.~Assuming that the required $\Delta n_{j}^{(p)}$ are achieved, we use Eq.~\reqt{stationary} to determine the ratio between the pump intensities in the first and second microrings as a function of the pump wavelength.
3.~At the pump wavelengths satisfying the amplitude ratio found in step 2, self consistency is achieved.
4.~We set the injected pump power according to the required intensities in the microrings. Note that the input pump intensity was irrelevant for the steps 2 and 3 where we analyzed a linear problem. This enables very efficient numerical calculation of the pump parameters.

\pict{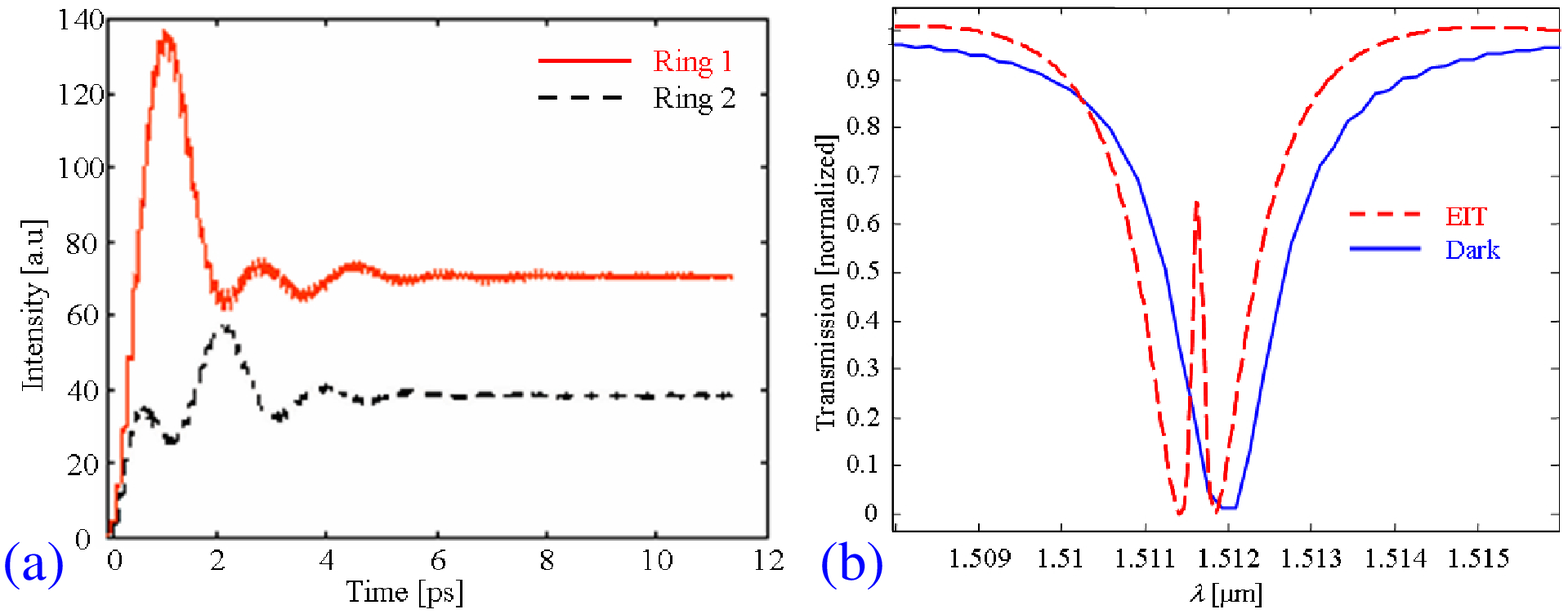}{dynamics}{
(Color online) Nonlinear FDTD simulation results:
(a)~Temporal evolution of the pump intensity in the first and second rings as indicated by labels;
(b)~Probe transmission spectra at $t=0$ (dashed line) and $t=11 ps$ (solid line).
}

To illustrate and validate our approach, we perform a comparison with direct modeling using the finite-difference-time-domain (FDTD) method. In order to reduce the computation time, small and high index-contrast microrings were employed. The dimensions of the simulated structure are $r_1 = r_2 = 1.7 \mu m$, $d=4.6421 \mu m$, ring/waveguide width $0.2 \mu m$, the gap between the waveguides and the rings is $w_{\rm gap} = 0.3 \mu m$, the refractive index of the waveguide and rings is $n_{\rm core}=3$, and the cladding refractive index is $n_{\rm clad}=1.0$. We use the linear transmission spectra and the values of phase accumulation in the rings and the waveguides obtained from the FDTD simulations to extract the matching parameters for our model: $\kappa = 0.0215$, $n_{10} = n_{20} = 2.2644$, $n_{0w} = 2.2796$. Under these conditions, the structure forms a ``dark state'' at the wavelength of $\lambda^{(s)} \simeq 1.5117 \mu m$ (with the resonant orders $m_1 = m_2 = 16$ and $m_w = 28$). To demonstrate the tunability between the "dark" and "EIT-like" states, the core (linear) indices of the microrings are slightly modified. The index of the first ring is set to $3.0-2 \delta n$ and that of the second ring to $3.0-\delta n$, where $\delta n=0.0005$. This approach is essentially identical to the modification of the radii but allows better control for computational process.
We determine the corresponding effective model parameters as $n_{10} = 2.2634$ are $n_{20} = 2.2639$. Substituting these parameters in Eq.~\reqt{stationary}, we immediately see the expected result that the resonances of two microrings are shifted apart and each resonance is associated with a dip in the transmission spectrum [Fig.~\rpict{transm}(a)] and the corresponding increase of the field intensity inside the microring [Fig.~\rpict{transm}(b)]. We apply the self-consistent approach and find several possible values of the pump wavelengths which enable us to achieve a ``dark state''. It is important to choose the pump wavelength sufficiently far from the microring resonance in order to avoid the bistability effect of the pump wave. In particular, we can choose the value of $\lambda^{(p)} = 1.7257 \mu m$. Then, to verify our assumption of equal intensities in left and right half-rings, we solve the fully nonlinear problem of the pump wave transmission by an iterative procedure: we calculate the pump intensities according to Eq.~\reqt{stationary}, then find the corresponding refractive index changes, and repeat these steps until convergence is reached. We use this solution to calculate the signal transmission spectrum shown in Fig.~\rpict{transm}(c), which indeed has a characteristic signature of a ``dark state'', where the two microting resonances are merged into one and the transmission is suppressed, whereas the field becomes much more strongly enhanced inside the microrings, c.f. Figs.~\rpict{transm}(d) and~(b).

We also characterize the temporal dynamics of the cavity tuning by performing FDTD modeling of nonlinear pump evolution. The numerically optimal value of the pump wavelength is found to be $\lambda^{(p)} = 1.7527$, which is close to the result of our semi-analytical approach. We observe a rapid convergence of the stationary amplitudes inside the microrings, within a few picoseconds [Fig.~\rpict{dynamics}(a)].
Figure~\rpict{dynamics}(b) depicts the small-signal transmission spectra of the coupled microrings system at $t=0$ and at $t=11ps$. As shown in the figure, when the pump is off, the device exhibits an EIT-like transmission spectrum with a sharp transmission line at $\lambda=1511.5 nm$. After the pump is turned on and the intensities in the microrings stabilize, a ``dark-state'' is formed and the compound cavity is ``closed''.

In conclusion, we have suggested a novel all-optical scheme for tuning a system of coupled nonlinear microring resonators
between "EIT-like" and "dark" states. These states correspond to opened and closed configurations of a compound cavity, thereby enabling the trapping
and releasing of optical pulses. The proposed scheme offers switching times of the order of few picoseconds being within the reach of
contemporary technology.

The authors thank the support of COST action MP0702 in this research. This work was partially supported by the Australian Research Council.

\end{document}